\documentclass[journal=nalefd, manuscript=letter, layout=traditional]{achemso}

\usepackage{siunitx}
\usepackage{chemformula}
\usepackage{amsmath}
\usepackage{amssymb}
\usepackage{graphicx}
\usepackage{hyperref}
\usepackage{verbatim}
\usepackage[latin1]{inputenc}
\setcitestyle{super}

\newcommand{\mr}[1]{\mathrm{#1}}

\newcommand{\cf}[0]{cf.\ }
\newcommand{\ie}[0]{i.e.\ }

\newcommand{\fref}[1]{Fig.~\ref{fig:#1}}

\newcommand{\eref}[1]{Eq.~(\ref{eq:#1})}
\newcommand{\Cref}[1]{Chapter~\ref{chap:#1}}
\newcommand{\cref}[1]{Ch.~\ref{chap:#1}}

\renewcommand{\vec}[1]{{\mathrm{\mathbf{#1}}}}

\newcommand{\MPS}[0]{\ch{Mn_{1.8}PtSn}}
\newcommand{\TSO}[0]{$T_\mr{SR}$}

\title{All Electrical Access to Topological Transport Features in \MPS\ Films}

\author{Richard Schlitz}
\email{richard.schlitz@tu-dresden.de}
\affiliation{Institut f{\"u}r Festk{\"o}rper- und Materialphysik, Technische Universit{\"a}t Dresden, 01062 Dresden, Germany}
\alsoaffiliation{Center for Transport and Devices of Emergent Materials, Technische Universit{\"a}t Dresden, 01062 Dresden, Germany}
\author{Peter Swekis}
\affiliation{Max-Planck Institute for Chemical Physics of Solids, 01187 Dresden, Germany}
\author{Anastasios Markou}
\affiliation{Max-Planck Institute for Chemical Physics of Solids, 01187 Dresden, Germany}
\author{Helena Reichlova}
\affiliation{Institut f{\"u}r Festk{\"o}rper- und Materialphysik, Technische Universit{\"a}t Dresden, 01062 Dresden, Germany}
\alsoaffiliation{Center for Transport and Devices of Emergent Materials, Technische Universit{\"a}t Dresden, 01062 Dresden, Germany}
\alsoaffiliation{Institute of Physics ASCR, v.\ v.\ i., Cukrovarnick{\'a} 10, 162 53, Praha 6, Czech Republic}
\author{Michaela Lammel}
\affiliation{Leibniz Institute for Solid State and Materials Research Dresden (IFW Dresden), Institute for Metallic Materials, 01069 Dresden, Germany}
\alsoaffiliation{Center for Transport and Devices of Emergent Materials, Technische Universit{\"a}t Dresden, 01062 Dresden, Germany}
\author{Jacob Gayles}
\affiliation{Max-Planck Institute for Chemical Physics of Solids, 01187 Dresden, Germany}
\author{Andy Thomas}
\affiliation{Leibniz Institute for Solid State and Materials Research Dresden (IFW Dresden), Institute for Metallic Materials, 01069 Dresden, Germany}
\alsoaffiliation{Center for Transport and Devices of Emergent Materials, Technische Universit{\"a}t Dresden, 01062 Dresden, Germany}
\author{Kornelius Nielsch}
\affiliation{Leibniz Institute for Solid State and Materials Research Dresden (IFW Dresden), Institute for Metallic Materials, 01069 Dresden, Germany}
\alsoaffiliation{Technische Universit{\"a}t Dresden, Institute of Materials Science, 01062 Dresden, Germany}
\author{Claudia Felser}
\affiliation{Max-Planck Institute for Chemical Physics of Solids, 01187 Dresden, Germany}
\author{Sebastian T. B. Goennenwein}
\affiliation{Institut f{\"u}r Festk{\"o}rper- und Materialphysik, Technische Universit{\"a}t Dresden, 01062 Dresden, Germany}
\alsoaffiliation{Center for Transport and Devices of Emergent Materials, Technische Universit{\"a}t Dresden, 01062 Dresden, Germany}
\date{\today}

\begin{document}

\begin{abstract}
	The presence of non-trivial magnetic topology can give rise to non-vanishing scalar spin chirality and consequently a topological Hall or Nernst effect.
	In turn, topological transport signals can serve as indicators for topological spin structures.
	This is particularly important in thin films or nanopatterned materials where the spin structure is not readily accessible.
	Conventionally, the topological response is determined by combining magnetotransport data with an independent magnetometry experiment. 
	This approach is prone to introduce measurement artifacts.
	In this study, we report the observation of large topological Hall and Nernst effects in micropatterned thin films of \MPS\ below the spin reorientation temperature \TSO$ \approx \SI{190}{\kelvin}$. 
	The magnitude of the topological Hall effect $\rho_\mr{xy}^\mr{T} = \SI{8}{\nano\ohm\meter}$ is close to the value reported in bulk \ch{Mn2PtSn}, and the topological Nernst effect $S_\mr{xy}^\mr{T} = \SI{115}{\nano\volt\per\kelvin}$ measured in the same microstructure has a similar magnitude as reported for bulk \ch{MnGe} ($S_\mr{xy}^\mr{T} \sim \SI{150}{\nano\volt\per\kelvin}$), the only other material where a topological Nernst was reported.
    We use our data as a model system to introduce a topological quantity, which allows to detect the presence of topological transport effects without the need for independent magnetometry data. 
    Our approach thus enables the study of topological transport also in nano-patterned materials without detrimental magnetization related limitations. 
	
\end{abstract}

Non-coplanar spin configurations, such as Skyrmions or other topological configurations are very exciting.\cite{Bogdanov:1989,Dzyaloshinsky:1958,Moriya:1960,Skyrme:1961,Roessler:2006,Muehlbauer:2009}
In magnetic materials, such topological spin structures can give rise to a finite scalar spin chirality $S_{i} \cdot (S_{j} \times S_{k})$ between neighbouring spins $S_{i,j,k}$ on the sites $i,j,k$.\cite{Taguchi:2001}
In a semi-classical picture, itinerant electrons moving through a spin texture with finite scalar spin chirality or a Skyrmion lattice\cite{Roessler:2006} accumulate a Berry phase, with the latter acting like an additional magnetic field on those electrons.\cite{Ye:1999, Tatara:2002, Taguchi:2001, Onoda:2004, Berry:1984} 
This (fictitious) magnetic field in turn leads to an additional contribution to the Hall and Nernst effects.\cite{Taguchi:2001, Neubauer:2009, Kanazawa:2011, Huang:2012, Shiomi:2013, Liu:2017, Liu:2018}
The connection between topological features in transport and non-trivial magnetic textures was experimentally demonstrated by comparison to Neutron diffraction results\cite{Neubauer:2009} or Lorentz transmission electron microscopy studies.\cite{Li:2013}
Thus, the presence of a topological Hall (THE) and/or Nernst (TNE) effect has been proposed as means to electrically detect the presence of Skyrmions or other topologically non-trivial spin textures in a multitude of materials.
Specifically, the THE was studied in many bulk \cite{Neubauer:2009, Lee:2009, Kanazawa:2011, Li:2017, Liu:2017} and thin film materials \cite{Huang:2012, Li:2013, Rana:2016, Liang:2015, Matsuno:2016, Ohuchi:2015, Gallagher:2017, Li:2018, Swekis:2018}, while until now the TNE was only investigated in bulk MnGe.\cite{Shiomi:2013}
Since the exact origin of the topological contribution is still vividly discussed, the topological Hall signal $\rho_\mr{xy}^\mr{T}$ is usually extracted using a heuristic method:
The magnetic field dependence of the Hall response 
\begin{equation}
    \rho_\mr{xy}(H) =  R_0 H + (S_\mathrm{A} \rho_\mr{xx}^2 + \alpha \rho_\mr{xx}) M(H) + \rho_\mr{xy}^\mr{T}
	\label{eq:toph_ex_intro}
\end{equation}
is linked to the magnetization $M(H)$, which is usually measured in a separate setup on unpatterned samples.
Here, $R_0$ is the ordinary Hall coefficient, $S_\mathrm{A}$ contains the instrinsic and side-jump contribution to the anomalous Hall effect (AHE), $\alpha$ is a measure for the skew scattering contribution to the AHE and $\rho_\mr{xx}$ is the resistivity of the material.\cite{Nagaosa:2013}

In this letter, we investigate the Hall and Nernst signals in \MPS\ thin films as a function of temperature and magnetic fields and report both a large THE and a large TNE comparable to the TNE observed in bulk \ch{MnGe}.\cite{Shiomi:2013}
\ch{Mn2PtSn} and its related compounds are known for hosting antiskyrmions\cite{Nayak:2017} (\ch{Mn_{1.4}Pt_{0.9}Pd_{0.1}Sn}) as well as having a large THE below a transition temperature \TSO\ in bulk \cite{Liu:2018} (\ch{Mn2PtSn}) and thin films\cite{Li:2018, Swekis:2018} (\ch{Mn_{1.5-2}PtSn}) and thus are promising candidates for studying topological transport features.
Motivated by these recent studies, we use \MPS\ as a model system to establish a technique enabling the extraction of the topological features without prior knowledge of the magnetization.
By combining the Hall and Nernst response measured in the same microstructure, we are able to remove the conventional (anomalous) transport features, revealing the salient features seen in the topological contributions.
Since our approach does not rely on magnetization measurements, it is viable also for nano-patterned materials.

\begin{figure}[th]
	\includegraphics[width=\columnwidth]{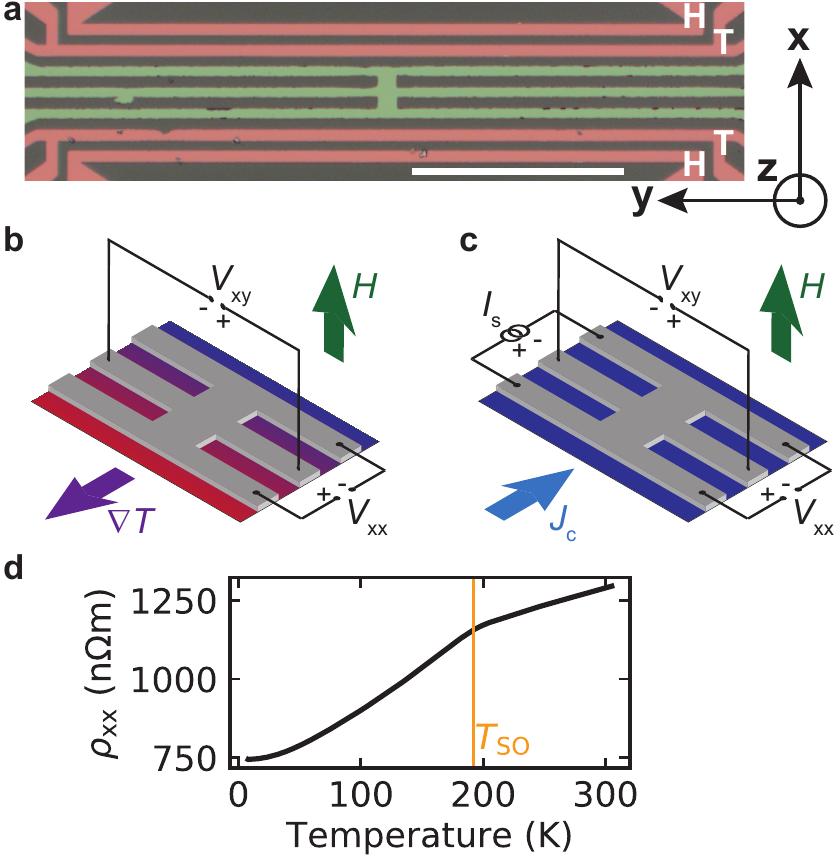}
	\caption{\label{fig:setup}
		\textbf{a} Optical micrograph of the sample. The white scalebar is \SI{100}{\micro\meter} long.
		The contacts used for the thermometry and application of the thermal gradients are colored in red and are made of platinum. The heaters and thermometers are marked with a white \textbf{H} and \textbf{T}, respectively.
		The \MPS\ micro Hall-bar and the contacts used for the measurement of the electric and thermal transport are highlighted in green.
		The coordinate system used for the discussion of the experiments is shown to the right.
		\textbf{b} Contact geometry for the Nernst and \textbf{c} contact geometry for the Hall measurements.
		The magnetic field is applied along the surface normal $\vec{z}$.
		A positive current is along the $\vec{x}$ direction, while a positive thermal gradient is anti-parallel to the $\vec{x}$ direction. 
		\textbf{d} The resistivity $\rho_\mr{xx}$ of the \MPS\ thin film decreases with decreasing temperature. 
		The spin reorientation temperature \TSO$\approx \SI{190}{\kelvin}$ is evident as a kink, as indicated by the orange line. 
	}
\end{figure}

\section{Methods}

The \MPS\ films were grown by DC magnetron co-sputtering of \ch{Mn}, \ch{Pt} and \ch{Sn} on MgO (001) substrates. The substrate temperature during deposition was \SI{350}{\celsius} and the Argon pressure was \SI{3e-3}{\milli\bar}. 
After deposition the samples were post-annealed for \SI{30}{\min} at the growth temperature and capped with a \SI{3}{\nano\meter} thick \ch{Al} capping layer at room temperature to prevent oxidation.
The crystalline quality was verified by X-ray diffraction and the thickness $t = \SI{35}{\nano\meter}$ was determined using X-ray reflectometry.
For further details about the film composition and structural properties refer to Ref. \cite{Swekis:2018}.

Magnetization loops were measured in a QuantumDesign MPMS XL7 on the as-grown samples and are shown in the supplementary information\cite{TNE-MPS-SI:2018}. 
For magnetotransport measurements the samples were patterned using optical lithography and subsequent \ch{Ar} ion milling.
In the last step, symmetric Pt heaters and thermometers were defined in a lift-off process with \SI{30}{\nano\meter} of sputtered Pt.
We will refer to the heaters and thermometers at one or the other end of the Hall-bar as ``top'' and ``bottom'' heater and thermometer, respectively.
An optical micrograph of a typical structure is shown in \fref{setup}\textbf{a}. 
All transport measurements were performed in a superconducting magnet cryostat with a magnetic field $|\mu_0 H_\mr{z}| \le \SI{6}{\tesla}$ applied along the $\vec{z}$ direction, perpendicular to the film plane.

To accurately measure the thermovoltage and thus obtain the Nernst signal, we employ an on-chip alternating gradient technique:
We alternatingly drive a current $I_\mr{H} = +\SI{5}{\milli\ampere}$ through the heaters at either side (\ie current through the bottom heater and no current through the top heater = heat flow direction $\uparrow$, and vice versa bottom heater off and current through the top heater = heat flow direction $\downarrow$) and simultaneously monitor the resistance $R_T$ of the two thermometers on top and bottom of the Hall-bar to determine the temperature gradient $\nabla T_\mr{x}$ along the $\vec{x}$ direction (MgO [100] axis).

To extract the part of the obtained voltage $V_\mr{xy}$ that is antisymmetric with respect to the gradient direction we subtract the two voltages measured for the two respective directions of heat flow $V_\mr{xy}(\uparrow)$ and $V_\mr{xy}(\downarrow)$ and calculate the Nernst signal as
\begin{equation}
    S_\mr{xy} = \frac{V_\mr{xy}(\uparrow)-V_\mr{xy}(\downarrow)}{2 l_\mr{H}} \frac{1}{\nabla T_\mr{x}}.
\end{equation}
Here, $l_\mr{H}$ is the length of the heaters, \ie the length of the contacts that is heated.
The inversion of the thermal gradient allows to remove spurious thermoeletric contributions caused by the setup in analogy to an electric current reversal technique or other modulation techniques applied in measuring a multitude of physical properties.\cite{Goennenwein:2015,Wu:1998, Kraftmakher:2002, Yagmur:2018} 
A sketch of the measurement principle is shown in \fref{setup}\textbf{b}.
For more information on the thermometry, see the supplementary information\cite{TNE-MPS-SI:2018}. 

To record the magnetoresistive and Hall response, we drive a current of $I_s = \SI{200}{\uA}$ along the Hall-bar with a Keithley 2450 sourcemeter.
The longitudinal and transverse voltage drop $V_\mr{xx}$ and $V_\mr{xy}$ are simultaneously detected by two Keithley 2182 nanovoltmeters. 
The measurement scheme is depicted in \fref{setup}\textbf{c}.
To increase the measurement sensitivity and to remove thermoelectric contributions to the voltage, we employ a current reversal technique.\cite{Goennenwein:2015}

The obtained temperature dependent resistivity $\rho_\mr{xx}$ of \MPS\ (\cf \fref{setup}\textbf{d}) shows a kink around \TSO$\approx \SI{190}{\kelvin}$, which we attribute to the spin reorientation already observed in bulk \ch{Mn2PtSn}\cite{Liu:2018} at $T_\mr{SR,bulk} = \SI{192}{\kelvin}$ and similarly in \ch{Mn2RhSn} at $T_\mr{SR,MRS} = \SI{80}{\kelvin}$.\cite{Meshcheriakova:2014}
Below \TSO, the magnetic sublattices change from a collinear to a non-collinear configuration.\cite{Bogdanov:2002, Swekis:2018}
 
\begin{figure}[h]
	\includegraphics[width=\columnwidth]{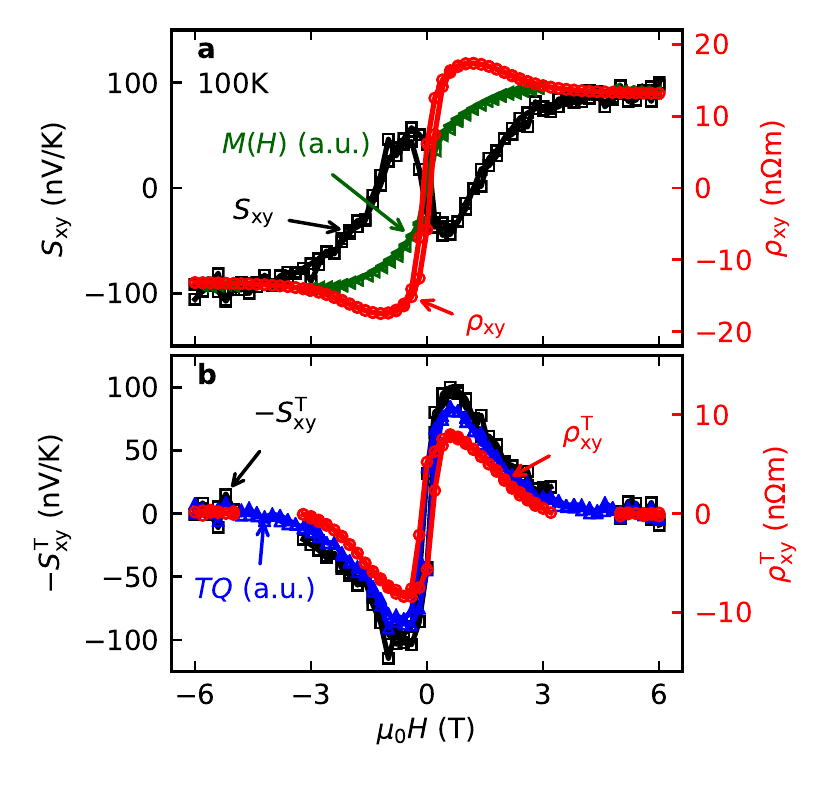}
	\caption{\label{fig:txe_eval}
	    \textbf{a} Nernst, Hall and magnetization measurements. 
	    Both transport signals have a different shape when compared to the magnetization.
	    \textbf{b} The topological quantity $TQ$ (blue triangles) reflects the magnetic field dependence of the TNE (black squares) and THE (red circles) and can be obtained without using the magnetometry data.
	}
\end{figure}

\section{Results and Discussion}

In the following, we will discuss the field dependence of the measured signals at $T=\SI{100}{\kelvin}$. 
This temperature is well enough below \TSO\ and simultaneously high enough to still observe a clear anomalous Nernst effect.
The Nernst and Hall effect signals, $S_\mr{xy}$ and $\rho_\mr{xy}$, respectively, are shown together with the magnetization in \fref{txe_eval}\textbf{a}.
In contrast to the curve shape expected for a ferromagnetic material -- where the signals mimic the magnetization -- an additional dip (peak) is visible in the Nernst (Hall) curve at positive magnetic fields (\cf\ \eref{toph_ex_intro}).\cite{Nagaosa:2013, Kanazawa:2011, Li:2013}
This additional feature is attributed to the aforementioned topological Nernst and Hall effect.\cite{Neubauer:2009, Shiomi:2013}
Both, the anomalous Nernst and Hall have the same sign in saturation, whereas the topological features are of opposite sign with respect to each other.
Note that in a free electron picture, the Hall and Nernst signals should be of opposite sign, since the electron velocity is in the opposite direction in the two measurement configurations (\cf \fref{setup}\textbf{b,c}).

To extract the topological transport response from the data, we now first follow the procedure customarily employed in the literature\cite{Nagaosa:2013, Swekis:2018, Shiomi:2013, Li:2018, Liu:2017} and subtract the anomalous effects from the measured curves using the magnetization loop in conjunction with a simplified version of \eref{toph_ex_intro}.
\begin{align}
	\label{eq:topn_ex}
	S_\mr{xy}^\mr{T}(H) = S_\mr{xy}(H) - S_\mr{xy}^\mr{A}\frac{M(H)}{M_\mr{s}} \\
	\rho_\mr{xy}^\mr{T}(H) = \rho_\mr{xy}(H) - \rho_\mr{xy}^\mr{A} \frac{M(H)}{M_\mr{s}} 
	\label{eq:toph_ex}
\end{align}
Here, $\rho_\mr{xy}^\mr{A}$, $S_\mr{xy}^\mr{A}$ and $M_\mr{s}$ are the amplitudes of the anomalous Hall and Nernst effect as well as the magnetization in the saturated field region ($\mu_0 H > \SI{4}{\tesla}$), respectively.
These equations neglect the ordinary Nernst and Hall effect signals as well as the contributions of the field dependent Seebeck and magnetoresistive signal $(S_\mr{xx}(H)/S_\mr{xx}(0))^2$ and $(\rho_\mr{xx}(H)/\rho_\mr{xx}(0))^2$, respectively.
Taking the magnetoresistance and the field dependent magneto-Seebeck signal to be smaller than \SI{10}{\percent}, the quantitative error using this simplified approach is approximately \SI{1}{\percent} when considering only intrinsic and side-jump scattering effects\cite{Swekis:2018}.
The resulting topological Nernst and Hall signals are shown in \fref{txe_eval}\textbf{b} with black and red symbols, respectively.
The missing points in the TNE and THE curves correspond to the zero-crossings in the $M(H)$ loops where the extraction of the magnetic moment fails (\cf supplementary material\cite{TNE-MPS-SI:2018}).

We now present a different approach to verify the presence of topological transport features. 
Building on the fact, that the anomalous Nernst and Hall signals both scale similarly with the magnetization, we can also take the difference of these two transport signals to remove the anomalous contributions instead of using the magnetization. 
Since the saturation values and units of the two effects are different, we hereby normalize both curves to their respective saturation values $\rho_\mr{xy}^\mr{A}$ or $S_\mr{xy}^\mr{A}$ and obtain
\begin{equation}
    \label{eq:tq_ex}
	TQ = \frac{\rho_\mr{xy}}{\rho_\mr{xy}^\mr{A}} - \frac{S_\mr{xy}}{S_\mr{xy}^\mr{A}} = \frac{\rho_\mr{xy}^\mr{T}}{\rho_\mr{xy}^\mr{A}} - \frac{S_\mr{xy}^\mr{T}}{S_\mr{xy}^\mr{A}}.
\end{equation}
As discussed above, the anomalous effects, scaling like the magnetization drop out entirely, leaving only the topological contributions.
The difference of the two normalized signals, which we name topological quantity $TQ$, is shown in \fref{txe_eval}\textbf{b} (blue triangles). 

The field dependence of the $TQ$, the TNE and the THE agree well, all having the same peak-dip structure in the intermediate field region ($\SI{-4}{\tesla} \le \mu_0 H \le \SI{4}{\tesla}$). 
This corroborates the idea, that the extraction of the topological features is possible using only electrical detection, even without the knowledge of the magnetization curves.
It is important to stress once more that the determination of the topological quantity does only require transport data, such that this approach is compatible with nano-patterned samples.
Eliminating the need for magnetization measurements furthermore allows to exclude several artifacts arising from comparing measurements taken in independent runs, in separates setups. 
In particular, as the Hall and Nernst signals are measured using the same contacts on the same sample in our approach, misalignment of the magnetic field with respect to the surface normal cannot contribute to the evaluation unless the sample is remounted between measurements.
Moreover, different temperature calibrations or local temperatures in different measurement setups can be ruled out.
The biggest caveat in using the $TQ$ extraction technique, is that the quantitative size of the TNE/THE is not known due to the normalization (\cf\ \eref{tq_ex}) and the possibly different contribution of the Nernst and Hall effects to the $TQ$.
However, the magnitude of $TQ$ still allows to infer an order of magnitude for the effect sizes.
The ranges for the maximum topological Nernst and Hall effect at a given temperature are determined by $S_\mr{xy}^T \le TQ\cdot S_\mr{xy}^\mr{A} \sim \SI{150}{\nano\volt\per\kelvin}$ and $\rho_\mr{xy}^T \le TQ\cdot \rho_\mr{xy}^\mr{A} \sim \SI{21}{\nano\ohm\meter}$, respectively. 
As expected, the observed maximum amplitudes of the TNE $S_\mr{xy}^T \sim \SI{100}{\nano\volt\per\kelvin}$ and THE $\rho_\mr{xy}^T \sim \SI{8}{\nano\ohm\meter}$ lie within these ranges. 
Taken together, the $TQ$ approach thus is a robust and scalable approach to infer the presence of topological transport features.
Further information on the extraction of the magnitudes and the extraction technique in general is given in the supplementary material.\cite{TNE-MPS-SI:2018}

\begin{figure}[h]
	\includegraphics[width=\columnwidth]{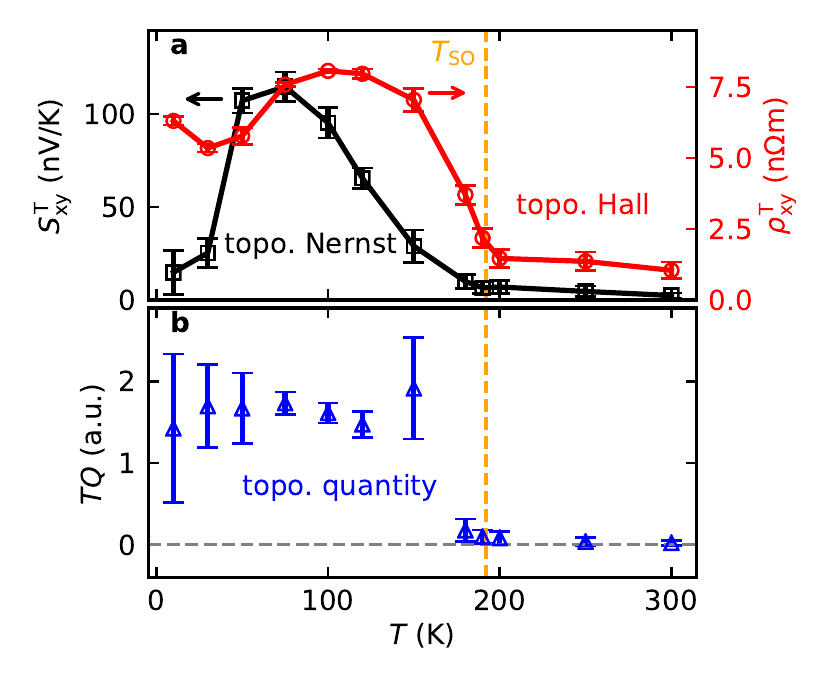}
	\caption{\label{fig:txe_t}
	    \textbf{a} The maximum amplitudes of the topological Nernst and Hall signals are shown by black and red symbols, respectively. 
		In contrast to the THE, the TNE decreases by at least a factor of 8 towards low temperatures ($T < \SI{50}{\kelvin}$).
		\textbf{b} The maximum of the topological quantity ($TQ$) decreases to $0$ within the experimental error above \TSO\ (marked by the orange line).
		Interestingly, the topological Hall and Nernst effect do not seem to completely vanish above \TSO.
	}
\end{figure}

Finally, we summarize the maximum amplitudes of the TNE and THE in \fref{txe_t}\textbf{a}.
The maximum TNE and THE over the full temperature range are $S_\mr{xy}^T(\SI{75}{\kelvin}) \sim\SI{115}{\nano\volt\per\kelvin})$ and $\rho_\mathrm{xy}^T(\SI{100}{\kelvin}) \sim \SI{8}{\nano\ohm\meter}$, respectively. 
Additionally, the maximum of $TQ$ is plotted in \fref{txe_t}\textbf{b}.
The topological quantity shows roughly the same temperature dependence as the THE with one significant difference:
Above the spin reorientation temperature marked as vertical orange line, the THE and TNE seem to remain finite, while the topological quantity decrease to zero within the measurement error. 
This suggests that indeed artifacts introduced by the comparison of the magnetometry and transport measurements in the conventional approach might be the origin of the features observed above \TSO.

In summary, we have presented measurements of the anomalous and topological Hall and Nernst effect in \MPS\ thin films, utilizing an alternating thermal gradient measurement technique. 
We observe clear topological contributions to both the Hall and Nernst response.
This is the first measurement of the topological Nernst in a thin film material, with a very large amplitude of $S_\mr{xy}^\mr{T}(\MPS) \approx \SI{115}{\nano\volt\per\kelvin}$ on-par with the value reported for bulk \ch{MnGe} $S_\mr{xy}^\mr{T}(\ch{MnGe}) \sim \SI{150}{\nano\volt\per\kelvin}$.\cite{Shiomi:2013} 

Furthermore, we demonstrated that by combining the Hall and Nernst signals measured in the same device, it is possible to detect the presence of topological features without using magnetization measurements.
The results open a pathway for an all-electric detection of topologically non-trivial magnetization textures in particular in micro- and nanostructures, where quantitative magnetometry experiments are very difficult and thus impede the conventional approach for the determination of the topological contributions.
As a side effect, the resulting topological quantity $TQ$ can be used as a complementary method to gauge artifacts introduced by the magnetization measurements: 
Although the TNE and THE extracted using the established method remain finite above the spin reorientation temperature, the $TQ$ vanishes.

\begin{acknowledgement}
We acknowledge fruitful discussions with J. K{\"u}bler and technical support by B. Weise as well as S. Piontek. Additionally, we acknowledge funding by the Deutsche Forschungsgemeinschaft (projects GO 944/4-2, SFB 1143/C08 and SPP 2137/403502666), by the Ministry of Education of the Czech Republic Grant No. LM2018110 and LNSM-LNSpin, Czech Science Foundation Grant No. 19-28375X, EU FET OpenRIA Grant No. 766566.
\end{acknowledgement}

\begin{suppinfo}
Details on the thermometry, the magnetometry data, the temperature evolution of the Hall and Nernst curves, the temperature dependent amplitudes of the anomalous Hall and Nernst as well as as discussion of the prerequisites of the extraction method are given in the supplementary information.
\end{suppinfo}

\bibliography{TNE_MnPtSn.bib}

\end{document}


\section{Thermometry and stability of the alternating thermal gradient}

Prior to measuring the Nernst signal, we calibrate the resistance of the Pt thermometers by applying different, constant and uniform temperatures to sample rod and thus the sample. 
To measure the resistance, we apply a current $I = \SI{100}{\micro\ampere}$.
Additionally, we use a current reversal technique to remove spurious thermoelectric contributions.\cite{Goennenwein:2015}
The result of this calibration for one of the thermometers is presented in \fref{sup_term}\textbf{a}.

We then use $R_T(T)$ of both Pt thermometers as calibration for the local temperatures above and below the structure.
The resulting temperature differences between the thermometers for the heat flowing upwards $\Delta T (\uparrow)$ (positive gradient) and downwards $\Delta T(\downarrow)$ (negative gradient) are used to estimate the thermal gradient along the structure 
\begin{equation}
\nabla T_\mr{x} = \frac{\Delta T (\uparrow) - \Delta T (\downarrow)}{2 d_\mr{therm}}.
\end{equation}
Here, $d_\mr{therm}$ is the separation of the thermometers along the heat gradient direction, \ie along $\vec{x}$. Please note, that a positive gradient is defined as being antiparallel to the $\vec{x}$-direction.

In a next step we apply a constant field of $\mu_0 H = \SI{2}{\tesla}$ along the film normal $\vec{n}$. 
Then, we vary the applied heating power and observe the difference in the temperature of the two thermometers for the two gradient directions (\cf \fref{sup_term}\textbf{b}).
The temperature difference agrees for the two gradient directions and increases linearly with the applied heating power. 
As expected, the simultaneously obtained Nernst voltage (shown as green symbols) also depends linearly on the power (or more precisely on the temperature difference).

Finally, to determine the field dependence of the thermometers, we measure the temperature during a field sweep.
The resulting temperatures are depicted in \fref{sup_term}\textbf{c}.
A weak dependence of the local temperatures on the magnetic field can be observed for both thermometers and gradient directions (most likely due to the ordinary magnetoresistance in Pt).
However, the visible field dependence in the temperature difference (\cf \fref{sup_term}\textbf{d}) is smaller than \SI{10}{\percent}. Only a small drift of the temperature difference with time can be observed.

\begin{figure}[h]
	\includegraphics[width=\columnwidth]{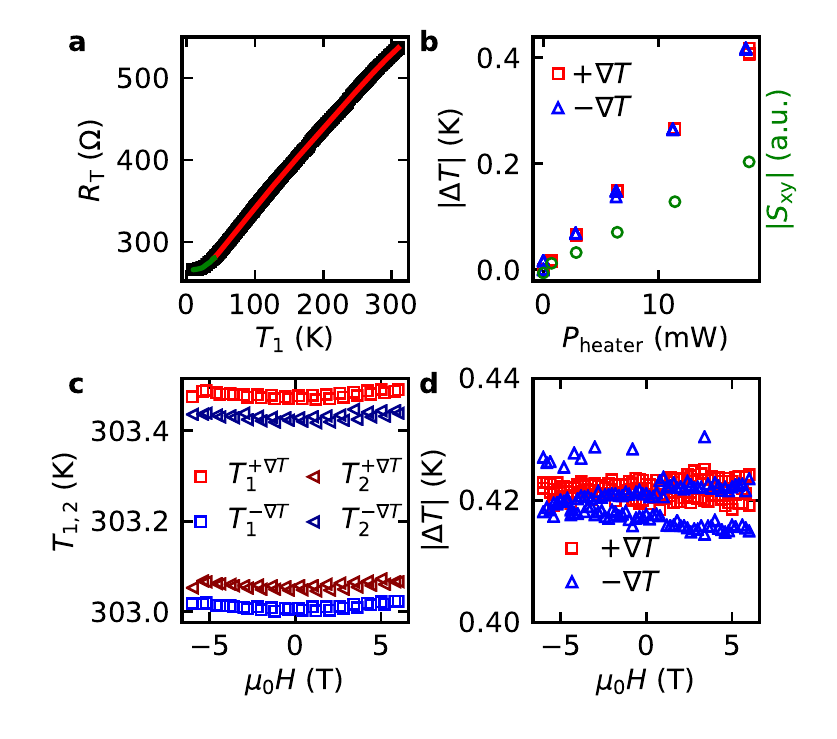}
	\caption{\label{fig:sup_term}
	\textbf{a} Calibration curve for one of the thermometers.
	The green and red line represent the low and high-temperature fit used for the calculation of the temperature.
	\textbf{b} The measured temperature differences for the two heat flow directions (blue and red symbols) as well as the Nernst signal (green circles) scale linearly with the applied heating power. 
	The measurements were taken with an applied magnetic field of $\mu_0 H_z = \SI{2}{\tesla}$.
	\textbf{c} The extracted temperatures on the hot and cold side recorded during one magnetic field sweep for the two heat flow directions are very similar and show only a small variation with the magnetic field. 
	\textbf{d} The extracted temperature difference is constant apart from a linear drift as a function of the applied magnetic field.
	}
\end{figure}

\section{Magnetization measurements}

The measurements of the out-of-plane magnetization were carried out in a QuantumDesign MPMS XL 7 SQUID magnetometer.
The obtained magnetization curves for all temperatures are shown in \fref{sup_mag}\textbf{a}. 
Due to the diamagnetic \ch{MgO} substrate contribution, the magnetization loop has additional zero-crossings at finite fields.
At these zero-crossings of the measured magnetization, the fit of the SQUID magnetometer fails, leading to additional measurement artifacts. 
Thus, the data around the zero-crossings have been masked using the fit quality.
To single out the magnetization of the film, the substrate contribution (\ie\ a negative slope) has to be subtracted from the data. 
The resulting magnetization of the film $M_\mathrm{film}$, which then is used for the extraction of the topological Hall and Nernst is shown in \fref{sup_mag}\textbf{b}.
Please note, that at low temperatures ($T \lesssim \SI{30}{\kelvin}$) there is an additional paramagnetic contribution due to impurities within the substrate. 
Since this letter focuses on the new extraction technique and its possible applications, we do not account for its effect here.
This will result in slightly different absolute values of the maximum topological Nernst and Hall contribution at these low temperatures.
For more details on the proper way to analyze this additional contribution please refer to Ref.\ \cite{Swekis:2018}.

\begin{figure}[h]
	\includegraphics[width=\columnwidth]{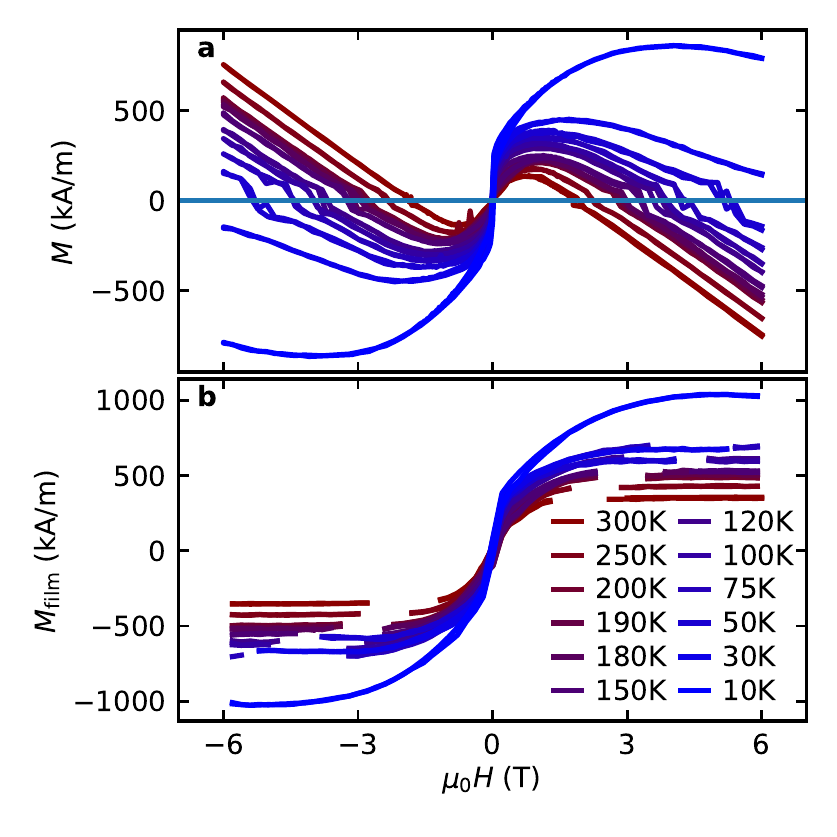}
	\caption{\label{fig:sup_mag}
	\textbf{a} The out-of-plane magnetization measurements of the \MPS\ thin film goes through zero at finite fields due to the diamagnetic substrate contribution, giving rise to measurement artifacts. 
	Therefore, the zero crossings have been masked.
	\textbf{b} Out-of-plane magnetization with subtracted background as used for the extraction of the topological Hall and Nernst effect.
	}
\end{figure}

\section{Temperature and field dependent measurements of the Hall and Nernst signal}

\begin{figure}[h]
	\includegraphics[width=\columnwidth]{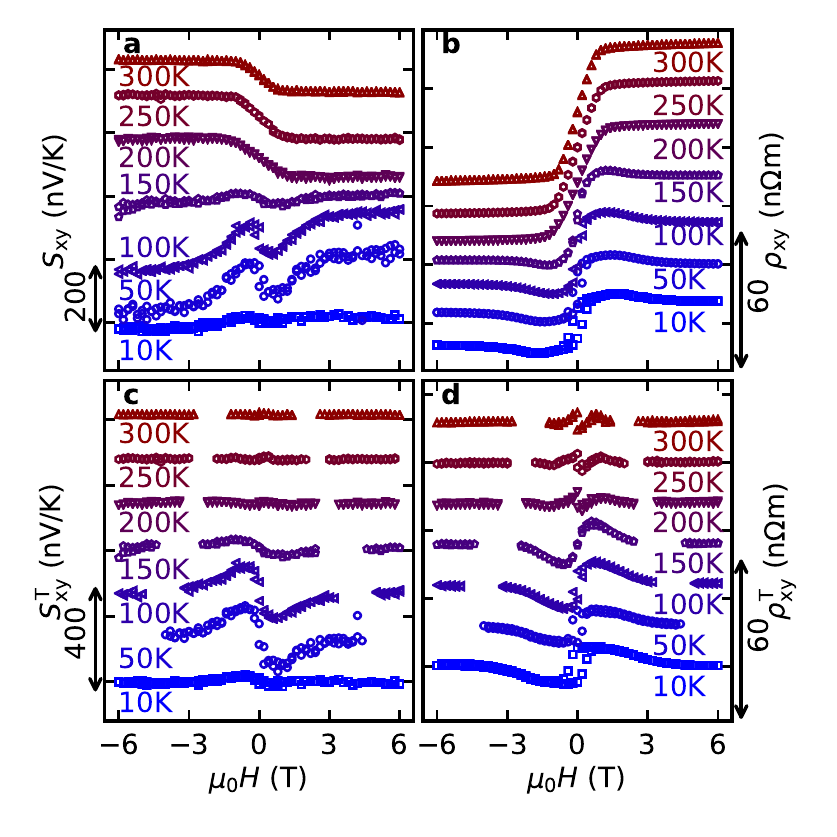}
	\caption{\label{fig:txe_ht}
		The evolution of the Nernst and Hall signal as a function of field and temperature is shown panels \textbf{a} and \textbf{b}, respectively.
		The anomalous Nernst effect (\ie the saturation value) changes sign below \TSO.
		The extracted topological Nernst and Hall signals are shown in \textbf{c} and \textbf{d}, respectively. 
		For $T > \SI{200}{\kelvin}$ no clear topological Hall and Nernst can be observed. In contrast, for $T < T_\mr{SR}$ and down to $T=\SI{10}{\kelvin}$, both, topological Hall and Nernst are present, indicating a non-coplanar spin texture in the film.
		At low temperature the Nernst signal decreases significantly as expected from the Mott relation.\cite{Xiao:2006, Xiao:2010}
	}
\end{figure}

The evolution of the Hall and Nernst response with temperature is shown in \fref{txe_ht}\textbf{a,b}, respectively.

When the temperature is below \SI{150}{\kelvin}, \ie below \TSO, silimar to those observed at \SI{100}{\kelvin} (see main text) are found. 
In particular, a dip-- or peak--like feature is present at low magnetic fields $\mu_0 H \lesssim \SI{4}{\tesla}$ in both Nernst and Hall signals and pertains to the lowest temperatures.
Additionally, when decreasing the temperature to below \SI{50}{\kelvin}, the Nernst signal decreases by roughly one order of magnitude.
This behaviour is expected for the anomalous thermal signals due to the Mott relation: Since the weighting function of the anomalous Nernst approaches a delta distribution, the anomalous Nernst signals will be suppressed towards $T=\SI{0}{\kelvin}$.\cite{Miyasato:2007, Noky:2018, Xiao:2006, Xiao:2010}

Please note, that the Nernst and Hall signals have opposite signs with respect to their anomalous counterparts below \TSO\ which was not observed in the only other report of the topological Hall and Nernst in the same sample.\cite{Shiomi:2013}

When increasing the temperature above the spin reorientation the obvious peak-- or dip--like feature visible below \TSO\ vanishes in both signals and the curves exhibit the behavior expected for conventional ferromagnets (\ie\eref{s_topn_ex} and \eref{s_toph_ex} with $S_\mr{xy}^{T} = 0$ and $\rho_\mr{xy}^{T} = 0$).

Additionally, the anomalous Nernst effect inverts its sign above \TSO.
In contrast, the anomalous Hall signal, although increasing for increasing temperatures, always exhibits the same sign.
We interpret this behaviour by considering the different contributing regions of the band structure:
While the anomalous Hall effect (AHE) is sensitive to all states below the Fermi surface, the anomalous Nernst effect (ANE) only probes a small region around the Fermi energy.\cite{Noky:2018, Xiao:2010}
Thus, a change of the band structure close to the Fermi energy will reflect differently in the ANE compared to the AHE.

The TNE and THE extracted from the measured curves using 
\begin{align}
	\label{eq:s_topn_ex}
	S_\mr{xy}^\mr{T}(H) = S_\mr{xy}(H) - S_\mr{xy}^\mr{A}\frac{M(H)}{M_\mr{s}} \\
	\rho_\mr{xy}^\mr{T}(H) = \rho_\mr{xy}(H) - \rho_\mr{xy}^\mr{A} \frac{M(H)}{M_\mr{s}} 
	\label{eq:s_toph_ex}
\end{align}
are shown in \fref{txe_ht}\textbf{c,d}.
Here, $\rho_\mr{xy}^\mr{A}$, $S_\mr{xy}^\mr{A}$ and $M_\mr{s}$ are the amplitudes of the anomalous Hall and Nernst effect as well as the magnetization in the saturated field region ($\mu_0 H > \SI{4}{\tesla}$), respectively. 
These equations neglect the ordinary Nernst and Hall, since their contributions are small compared to the anomalous counterparts.
Both topological signals pertain down to the lowest measured temperature $T = \SI{10}{\kelvin}$ extending from very small fields until roughly \SI{4}{\tesla}.
Interestingly, even above \TSO\ small features remain visible in the Hall and Nernst signal.
This observation would agree with the reported occurrence of Antiskyrmions above \TSO\ in bulk \ch{Mn_{1.4}Pt_{0.9}Pd_{0.1}Sn}.\cite{Nayak:2017}
However, a small misalignment or different temperatures of the sample during the transport or the magnetization measurements can give rise to slightly different $M(H)$ and $\rho_\mr{xy}(H)$ viz. $S_\mr{xy}(H)$ shapes and thus artifacts having a similar shape as the topological features.
The regions where no data for the TNE and THE are shown correspond to the masked regions in the $M(H)$ loops as discussed in the previous section.

\section{Anomalous Hall and Nernst effect}

We show the amplitudes of the ANE $S_\mr{xy}^\mr{A}$ and AHE $\rho_\mr{xy}^\mr{A}$ as a function of the temperature in \fref{sup_axet}.
As discussed above, the ANE (black symbols) changes its sign around \SI{150}{\kelvin} while the AHE only increases monotonically from low to high temperatures.
The maximum amplitudes are $S_\mr{xy}^\mr{A} \approx \SI{110}{\nano\volt\per\kelvin}$ and $\rho_\mr{xy}^\mr{A} \approx \SI{29}{\nano\ohm\meter}$.

\begin{figure}[h!]
	\includegraphics[width=\columnwidth]{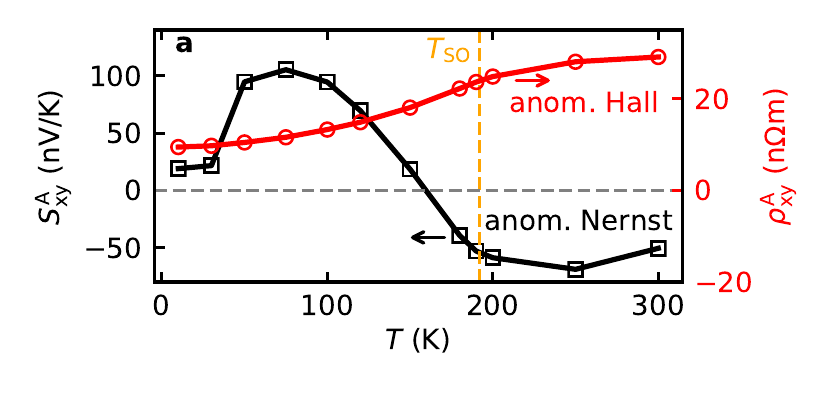}
	\caption{\label{fig:sup_axet}
	    \textbf{a} The saturation values of the Nernst (black squares) and Hall (red circles) signals, \ie the anomalous Hall and Nernst signals both show a dependence on temperature: While the AHE only decreases towards lower temperatures, the ANE changes its sign below the spin reorientation temperature. 
	}
\end{figure}

\section{Determination of the topological quantity and prerequisites for the methodology}

To extract the topological quantity we normalize the Nernst and Hall signals to their saturation value in the field polarized configuration (\ie where the magnetization texture and therefore the topological contribution is not present) and take the difference of both signals.
Thus, together with \eref{s_topn_ex} and \eref{s_toph_ex}, we get the following relation, again without considering the ordinary effects:
\begin{align}
	TQ = \frac{\rho_\mr{xy}}{\rho_\mr{xy}^\mr{A}} - \frac{S_\mr{xy}}{S_\mr{xy}^\mr{A}} = \frac{\rho_\mr{xy}^\mr{T}}{\rho_\mr{xy}^\mr{A}} + \frac{M(H)}{M_\mr{s}} -\left( \frac{S_\mr{xy}^\mr{T}}{S_\mr{xy}^\mr{A}} + \frac{M(H)}{M_\mr{s}}\right)= \frac{\rho_\mr{xy}^\mr{T}}{\rho_\mr{xy}^\mr{A}} - \frac{S_\mr{xy}^\mr{T}}{S_\mr{xy}^\mr{A}} = \widetilde{\rho_\mr{xy}^\mr{T}} - \widetilde{S_\mr{xy}^\mr{T}}
\end{align}
To also account for the ordinary effects, they would need to be subtracted prior to the subtraction, by using a linear fit in the field polarized phase.
From this result we infer, that as long as the ratio of the topological and the anomalous Hall and Nernst signals are not exactly the same, our method will yield usable results.
Since all signals do change significantly with temperature -- having vastly different temperature dependencies -- for the material in this study, the method should be robust.

Considering all signs as observed in this manuscript, we see that at low temperature the most advantageous case is realized, \ie\ where the two topological signals add. 
This can be understood as the topological and anomalous Nernst are both positive, while the anomalous and topological Nernst have opposite sign (\ie $TQ =  |\widetilde{\rho_\mr{xy}^\mr{T}}| + |\widetilde{S_\mr{xy}^T}|$).
However, at high temperatures ($T > T_\mr{SR}$) the topological and Nernst have the same sign, thus the two fractions are subtracted (\ie $TQ = |\widetilde{\rho_\mr{xy}^\mr{T}}| - |\widetilde{S_\mr{xy}^T}|$).
The only case where our method fails is if either the anomalous Hall or Nernst vanishes (as observed in this manuscript at the sign change close to \TSO). 
In this case the signal would be normalized by 0, such that the extraction method fails.

Although no exact numbers for the topological Hall and Nernst can be extracted using this method, a range for the two effect magnitudes can be given:
For example, lets consider that there is only a topological Hall effect ($S_\mr{xy}^\mr{T} = 0$). 
In this case, we can see that the amplitude of the THE is given by $\rho_\mr{xy}^\mr{T} = TQ\cdot\rho_\mr{xy}^\mr{A}$.
In a similar fashion, if there would be only a topological Nernst effect, its amplitude would be given by $S_\mr{xy}^\mr{T} = - TQ\cdot S_\mr{xy}^\mr{A}$.
If both transport signals contribute to the $TQ$, then these two values define the upper limit for the two effects, \ie\ $|S_\mr{xy}^\mr{T}| \le |TQ\cdot S_\mr{xy}^\mr{A}|$ and $|\rho_\mr{xy}^\mr{T}| \le |TQ\cdot\rho_\mr{xy}^\mr{A}|$. 

\section{Validity of the extraction method for other materials}

To further show the validity of our approach, we extracted the published data for the Hall and Nernst in bulk \ch{MnGe} from Ref. \cite{Shiomi:2013} for $T = \SI{20}{\kelvin}$ and $T = \SI{100}{\kelvin}$ (\cf \fref{sup_mnge}\textbf{a,b}).
We then apply the same approach as for our \MPS\ thin films, normalizing the signals and subtracting the normalized Hall and Nernst signals.
The resulting curves for the topological Hall and Nernst as well as the topological quantity are shown in \fref{sup_mnge}\textbf{c,d}.
\begin{figure}[h]
	\includegraphics[width=\columnwidth]{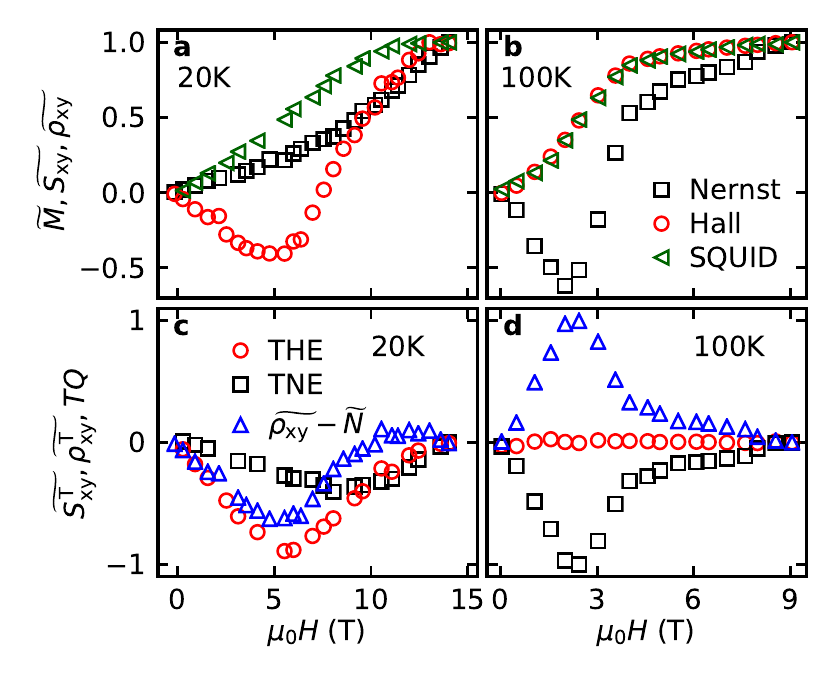}
	\caption{\label{fig:sup_mnge}
	\textbf{a,b} Normalized Nernst $\widetilde{S_\mr{xy}}=S_\mr{xy}/\max(S_\mr{xy})$, Hall $\widetilde{\rho_\mr{xy}}=\rho_\mr{xy}/\max(\rho_\mr{xy})$ and SQUID magnetometry $\widetilde{M} = M/\max(M)$ measurements for bulk \ch{MnGe} as reported in Ref. \cite{Shiomi:2013}.
	\textbf{c,d} The extracted topological quantity $TQ$ (blue triangles) also for this sample reflects the region where either topological Hall or Nernst are observed.
	}
\end{figure}
Again it can be seen, that although the THE and TNE are significantly different for the two temperatures, the $TQ$ is finite for the regions where either THE or TNE is present as expected from our preliminary discussion.
Please note, that here, for both temperatures the anomalous Hall and Nernst as well as the anomalous Hall have the same shape (\ie a dip--like structure when increasing the field).
This shows that even in the ``unfavorable'' case, where the topological effects do not add, our approach is valid.

\bibliography{TNE_MnPtSn}